\documentclass[11pt]{article}
\usepackage{moriond,epsfig}

\def\Journal#1#2#3#4{{#1} {\bf #2}, #3 (#4)}

\def\PRD{{\em Phys. Rev.} D}

\def\be{\begin{equation}}
\def\ee{\end{equation}}
\def\bea{\begin{eqnarray}}
\def\eea{\end{eqnarray}}

\begin{document}

\vspace*{4cm}
\title{FIRST RUN II MEASUREMENT OF THE W BOSON MASS WITH CDF}

%\author{ OLIVER STELZER-CHILTON\footnote{on behalf of the \uppercase{CDF} collaboration} }
\author{OLIVER STELZER-CHILTON for the CDF Collaboration}

\address{University of Oxford, Dept. of Physics, Denys Wilkinson Building,\\
Keble Road, OX1 3RH, Oxford, United Kingdom}

\maketitle\abstracts{
The CDF collaboration has analyzed $\sim$200 pb$^{-1}$ of Tevatron Run II data taken
between February 2002 and September 2003 to measure the $W$ boson mass. 
With a sample of 63964 $W\rightarrow e\nu$ decays and 51128 $W\rightarrow\mu\nu$ decays, 
we measure $M_W$ = 80413$\pm$34(stat)$\pm$34(syst) MeV/$c^2$. The total measurement 
uncertainty of 48 MeV/$c^2$ makes this result the most precise single measurement 
of the $W$ boson mass to date.}

\section{Introduction}
The $W$ boson mass is an important Standard Model (SM) parameter. It receives self-energy corrections due to 
vacuum fluctuations involving virtual particles. Thus the $W$ boson 
mass probes the particle spectrum in nature, including particles that have yet 
to be observed directly. The hypothetical particle of most immediate interest is the 
Higgs boson. The interaction of the other SM particles, in particular the $W$ and $Z$
gauge bosons, with the Higgs field is thought to impart mass to the SM particles. 
Thus the Higgs boson plays a critical 
role in the SM and it is very interesting to obtain a Higgs mass prediction.
The $W$ boson mass can be calculated at tree level using
the three precise measurements of the $Z$ boson mass, the Fermi coupling $G_F$ and the 
electromagnetic coupling $\alpha_{em}$. In order to extract information 
on new particles, we need to account for the radiative corrections to $M_W$ due to the dominant top-bottom 
quark loop diagrams. For fixed values of 
other inputs, the current uncertainty on the top quark mass measurement 170.9$\pm$1.8 GeV/$c^2$ \cite{top} corresponds to an 
uncertainty in its $W$ boson mass correction of 11 MeV/$c^2$. Measurements of the $W$ boson mass from Run I 
of the Tevatron and LEP with uncertainties of 59 MeV/$c^2$ \cite{tevwmass} and 33 MeV/$c^2$ \cite{lepwmass} respectively, yield a world 
average of 80392$\pm$29 MeV/$c^2$ \cite{lepwmass}. It is clearly profitable to reduce the $W$ boson mass uncertainty further 
as a means of constraining the Higgs boson mass.

\section{Measurement Strategy}
At the Tevatron, $W$ bosons are mainly produced by valance quark-antiquark annihilation, with initial
state gluon radiation (ISR) generating a typical transverse boost of O(10 GeV). The transverse momentum ($p_T^l$) distribution 
of the decay lepton has a characteristic Jacobian edge whose location, while sensitive to the $W$ boson mass, 
is smeared by the transverse boost of the $W$ boson. The neutrino transverse momentum ($p_T^{\nu}$) can be inferred by imposing $p_T$ 
balance in the event. The transverse mass, defined as $m_T=\sqrt{2 p_T^l p_T^{\nu}(1-cos[\phi^l-\phi^{\nu}])}$, includes both 
measurable quantities in the $W$ decay and provides the most precise quantity to measure the $W$ boson mass. 
We use the $m_T$, $p_T^l$ and $p_T^{\nu}$ distributions to extract $M_W$. These distributions do not 
lend themselves to analytic parameterizations, which leads us to use a Monte Carlo simulation to predict 
their shape as a function of $M_W$. These lineshape predictions depend on a number
of physical and detector effects, which we constrain from control samples or
calculation. Important physical effects include internal QED radiation, the intrinsic $W$ boson 
transverse momentum, and the proton parton distribution functions. Detector effects include external
bremsstrahlung and ionization energy loss in the detector material, tracker momentum scale,
calorimeter energy scale, resolution of the tracker and calorimeter, and the detector fiducial
acceptance. In order to model and study these effects at the level of a part in 10$^4$, we have
developed a sophisticated, first-principles fast Monte Carlo simulation of the CDF detector. 
The $W$ boson mass is extracted by performing a binned maximum-likelihood fit to the $m_T$, $p_T^l$ 
and $p_T^{\nu}$ distributions. We generate 800 templates as a function of $M_W$ between 80 GeV/$c^2$
and 81 GeV/$c^2$.

\section{Momentum and Energy Scale Calibration}
\begin{figure}[ht]
%=\epsfxsize=10cm   %width of figure - will enlarge/reduce the figures
%\epsfbox{fig3.eps}
%\figurebox{2cm}{3cm}{} %to have a box alone 
\centerline{\epsfxsize=3.3in\epsfbox{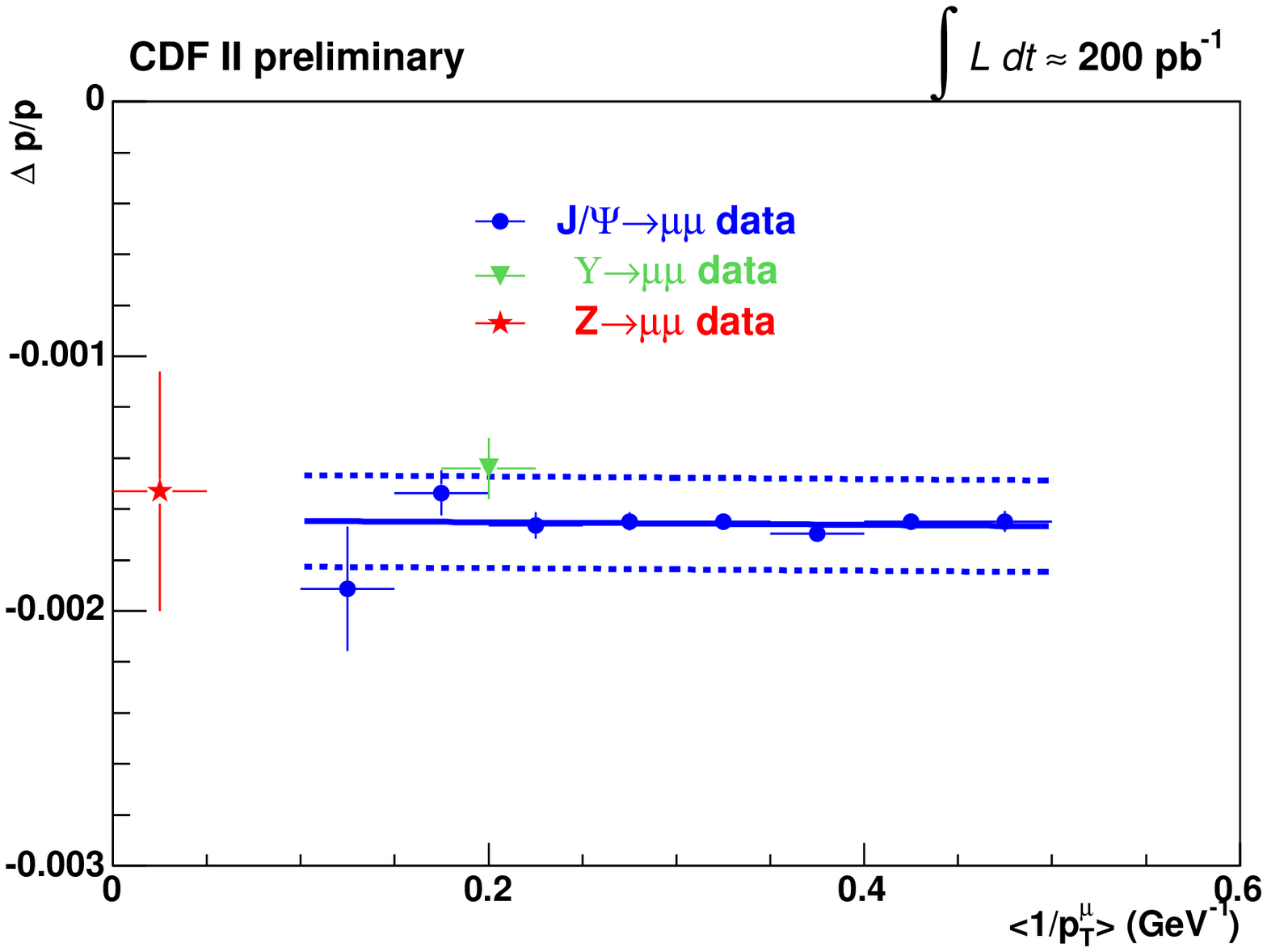}\epsfxsize=3.3in\epsfbox{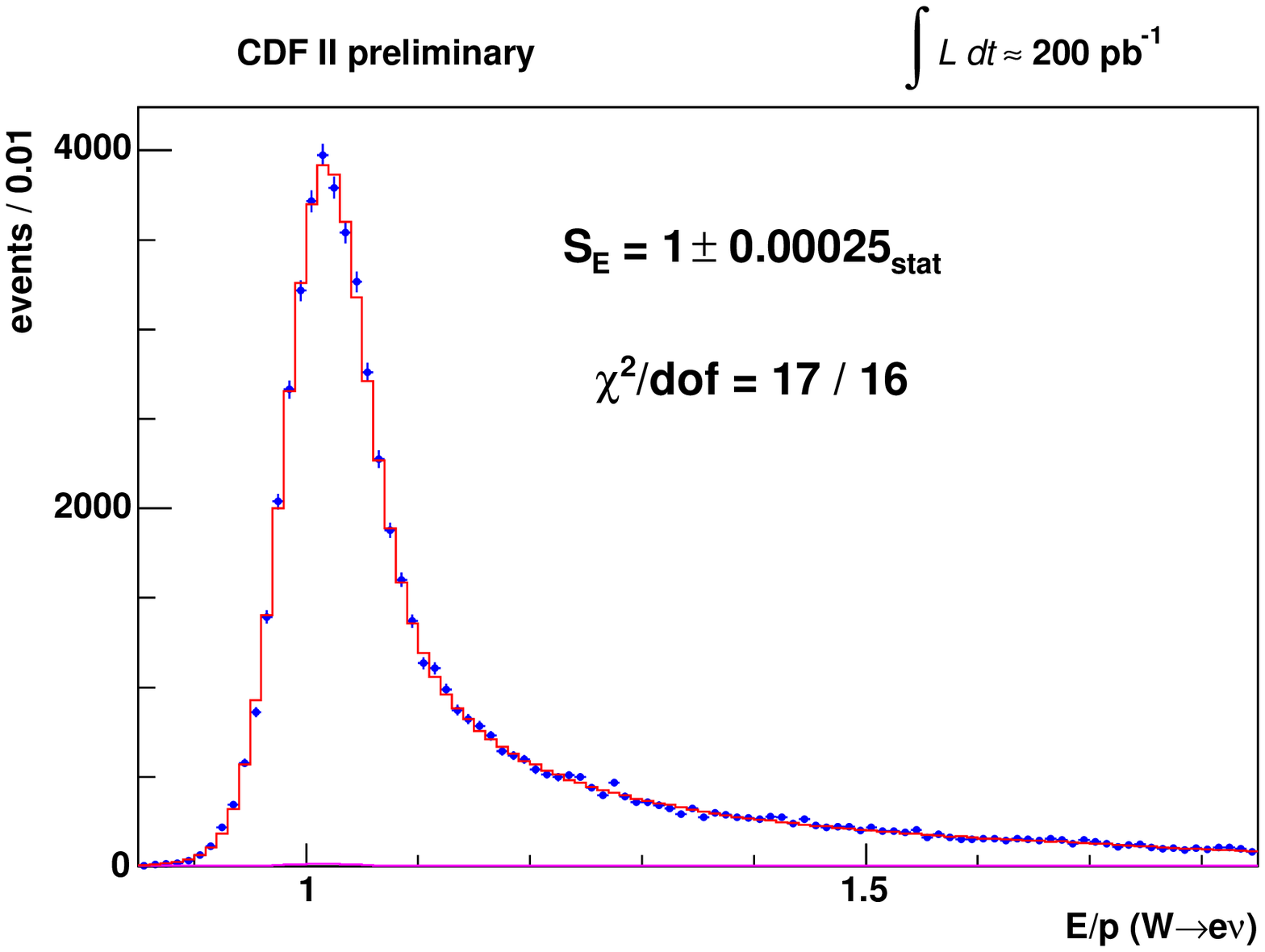}}
\caption{Left: Momentum scale summary: $\Delta p/p$ vs $1/p_T$ for $J/\Psi$, $\Upsilon(1S)$ and $Z$ boson dimuon samples. The dotted
line represents the independent uncertainty between J/$\Psi$ and $\Upsilon(1S)$.
Error bars are statistical only. \newline Right: Energy scale calibration
using $E/p$ distribution from $W\rightarrow e\nu$ events. \label{scales}}
\end{figure}

The key aspect of the measurement is the calibration of the lepton momentum. The charged
lepton momentum is measured in a cylindrical drift chamber called the Central Outer Tracker (COT). 
The electron energy is measured using the central electromagnetic (EM) calorimeter and its angle measurement
is provided by the COT trajectory. The COT track measurement sets the momentum scale for this analysis. The
internal alignment of the COT is performed using high-$p_T$ cosmic rays that traverse diametrically the entire
drift chamber. 
The momentum scale is set by measuring the $J/\Psi$ and $\Upsilon(1S)$ 
masses using the dimuon mass peaks. The $J/\Psi$ sample spans a range of muon $p_T$ (2-10 GeV/c), which allows us 
to tune our ionization energy loss model such that the measured mass is
independent of muon $p_T$. We obtain consistent 
calibrations from the $J/\Psi$, $\Upsilon(1S)$ mass fits shown in Fig. \ref{scales} (left). 
The momentum scale extracted from the $Z\rightarrow\mu\mu$ mass fit, shown in the same figure, 
is also consistent, albeit with a larger, statistics-dominated uncertainty.
The tracker resolution is tuned
on the observed width of the $\Upsilon(1S)$ and $Z$ boson mass peaks.

Given the tracker momentum calibration, we fit the peak of the
$E/p$ distribution of the signal electrons in the $W\rightarrow e\nu$ sample (Fig.~\ref{scales} right)
in order to calibrate the energy measurement of the electromagnetic (EM) calorimeter. The energy scale 
is adjusted such that the fit to the peak returns unity. 
The model for radiative energy loss is constrained, by comparing the number of events in 
the radiative tail of the $E/p$ distribution. The calorimeter
energy calibration is performed in bins of electron $E_T$ to constrain the calorimeter non-linearity.
The calibration yields a $Z\rightarrow ee$ mass measurement of $M_Z = 91190\pm67_{stat}$ MeV/$c^2$, in very
good agreement with the world average ($91187.6\pm2.1$ MeV/$c^2$ \cite{lepwmass}); we obtain the most precise calorimeter calibration by combining the
results from the $E/p$ method and the $Z\rightarrow ee$ mass measurement. The EM calorimeter resolution model
is tuned on the widths of the $E/p$ peak and the $Z\rightarrow ee$ mass peak, separately for non-radiative and radiative
electrons.

\section{Hadronic Recoil Calibration}
\begin{figure}[h]
%=\epsfxsize=10cm   %width of figure - will enlarge/reduce the figures
%\epsfbox{fig3.eps}
%\figurebox{2cm}{3cm}{} %to have a box alone 
\centerline{\epsfxsize=3.3in\epsfbox{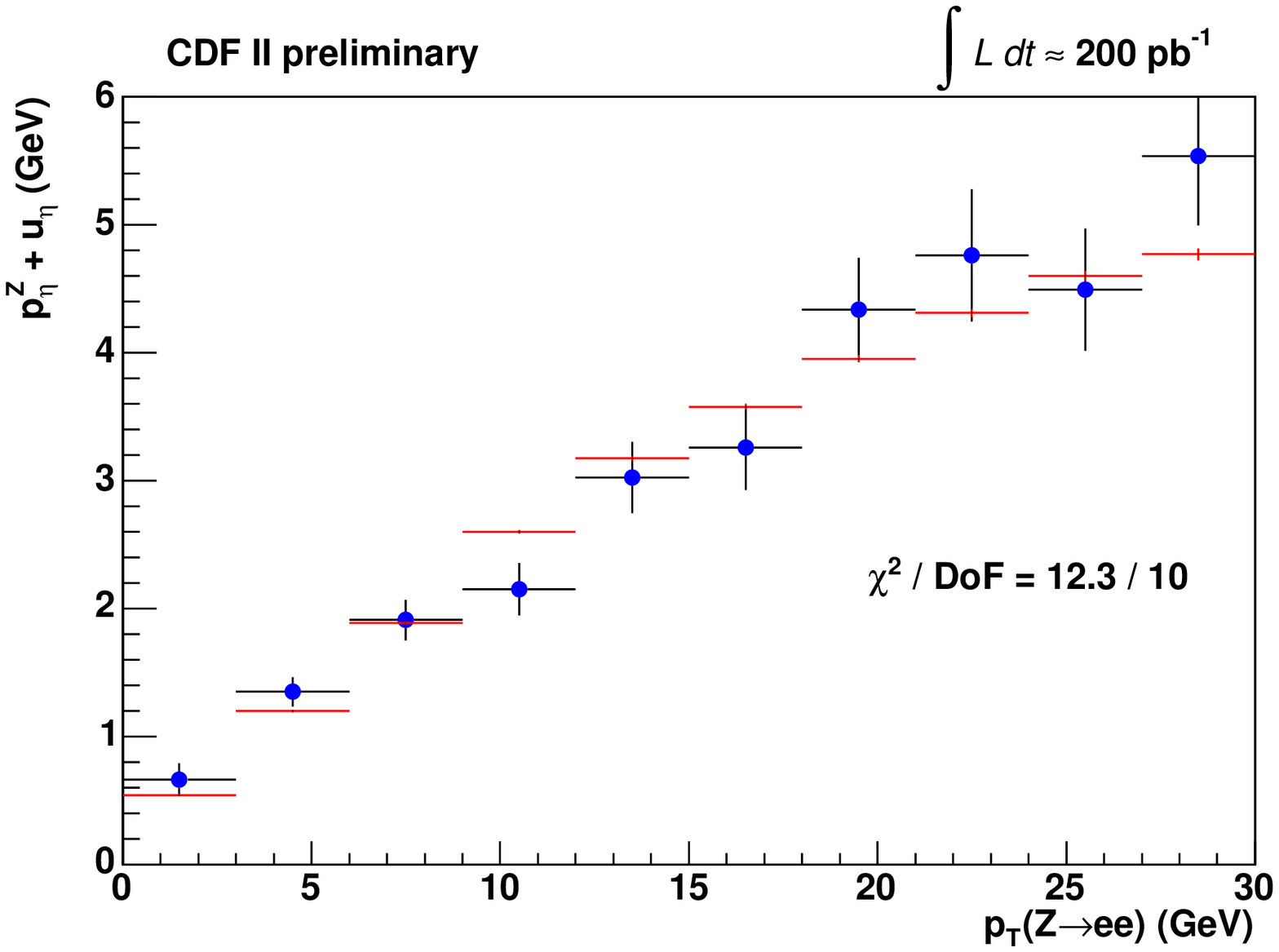}\epsfxsize=3.3in\epsfbox{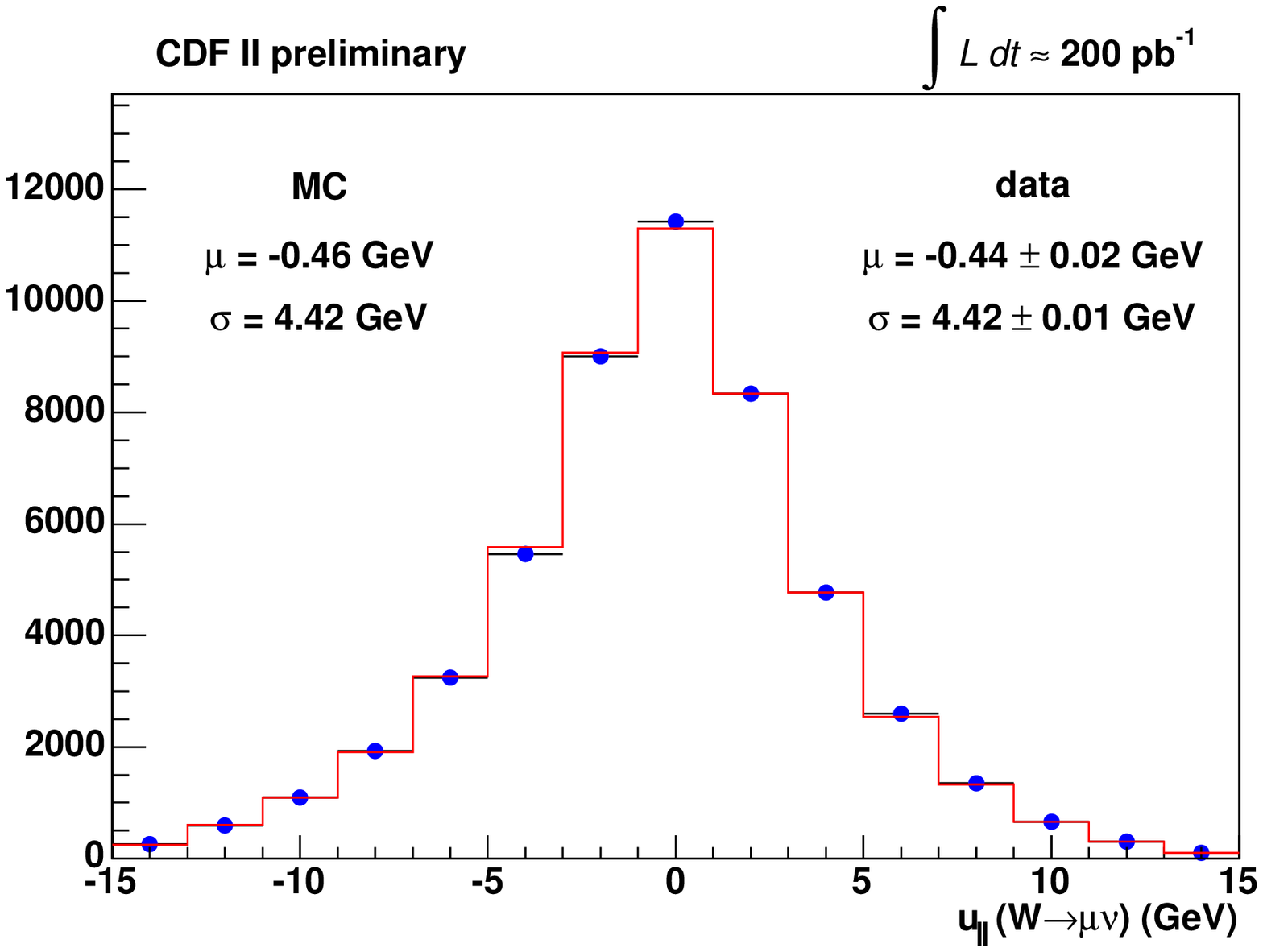}}
\caption{Left: Mean of the $p_T$-balancing as a function of $p_T$ in $Z\rightarrow ee$ events. Right: Projection of
recoil along the lepton direction in $W\rightarrow \mu\nu$ events. \label{recoil}}
\end{figure}

All particles recoiling against the $W$ or $Z$ boson are collectively referred to as the recoil.
The recoil is computed as the vector sum of transverse energy over all calorimeter towers, where the towers
associated with the leptons are explicitly removed from the calculation. The response of the calorimeter
to the recoil is described by a response function which scales the true recoil magnitude to simulate the
measured magnitude. The hadronic resolution receives contributions from ISR jets and the underlying event. The
latter is independent of the boson transverse momentum and modeled using minimum bias data.
The recoil response and resolution parameterizations are tuned on the mean and $rms$ of the
$p_T$-imbalance in $Z\rightarrow ll$ events as a function of boson $p_T$. We define the $\eta$ axis to be
the geometric bisector of the two leptons and the $\xi$ axis to be perpendicular to $\eta$. We
project the vector $p_T$-balance onto the $\eta$ and $\xi$ axes and compare the data distribution
to the simulation. Fig. \ref{recoil} (left) shows the mean of the $p_T$-balancing in $Z\rightarrow ee$ events
as a function of $Z$ boson $p_T$. 
The quantity $p_{\eta}^Z$ is computed from the EM clusters of the two $Z$ 
boson decay electrons and $u_{\eta}$ is computed from the recoil vector in the calorimeter.
We cross-check the recoil model using $W$ and $Z$ boson data which show good agreement
and validate the model. A very sensitive quantity to cross-check the recoil model in $W$ boson events is the projection of 
the recoil along the lepton direction ($u_{||}$). Fig. \ref{recoil} (right) shows $u_{||}$
for $W\rightarrow \mu\nu$ events. We find good agreement between the data and the simulation.

\section{Event Generation}
We generate $W$ and $Z$ events with {\sc resbos} \cite{resbos}, which captures the QCD physics and
models the $W$ $p_T$ spectrum. {\sc resbos} calculates the quintuple differential production cross section
$\frac{d^5\sigma}{dQ\,dy\,dq_T\,d^2\Omega}$, where $Q$, y and $q_T$ are the boson invariant mass, rapidity,
and transverse momentum respectively, and $d\Omega$ is the solid angle element in the decay lepton direction.
The {\sc resbos} parametrization of the
non-pertubative form factor is tuned on the dilepton $p_T$ distribution in the $Z$ boson
sample. Photons radiated off the final-state leptons (FSR) are generated according to
{\sc wgrad} \cite{wgrad}. The FSR photon energies are increased by 10\% (with an absolute uncertainty of 5\%) to
account for additional energy loss due to two-photon radiation \cite{2photon}. {\sc wgrad} is also used
to estimate the uncertainty due to QED radiation from initial state (ISR) and interference between ISR and FSR.
We use the CTEQ6M \cite{cteq} set of parton distribution 
functions at NLO and apply their uncertainties to evaluate the systematic uncertainty on the $W$
boson mass. 

\section{Backgrounds}
Backgrounds passing the event selection have different kinematic distributions from the $W$ signal and are
included in the template fit according to their normalizations. Backgrounds arise in the $W$ boson samples 
from misidentified jets containing high-$p_T$ tracks and
EM clusters, $Z\rightarrow ll$ where one of the leptons is not reconstructed and mimics a neutrino,
$W\rightarrow\tau\nu$, kaon and pion decays in flight (DIF), and cosmic ray muons.
The latter two are backgrounds in the muon channel only. Jet, DIF, and cosmic ray backgrounds are estimated from the 
data to be together less than 0.5\%. The $W\rightarrow\tau\nu$ background is 0.9\% for both channels, and 
the $Z\rightarrow ll$ is 6.6\% (0.24\%) in the muon (electron) channel, as estimated from Monte Carlo samples
generated with {\sc pythia} \cite{pythia} and a detailed {\sc geant}-based detector simulation.

\section{Results and Conclusions}
The fits to the three kinematic distributions $m_T$, $p_T^l$ and $p_T^{\nu}$ in the electron and muon channels give
the $W$ boson mass results shown in Table \ref{fits}.

\begin{figure}[ht]
%=\epsfxsize=5cm   %width of figure - will enlarge/reduce the figures
%\epsfbox{fig3.eps}
%\figurebox{2cm}{3cm}{} %to have a box alone 
\centerline{\epsfxsize=3.3in\epsfbox{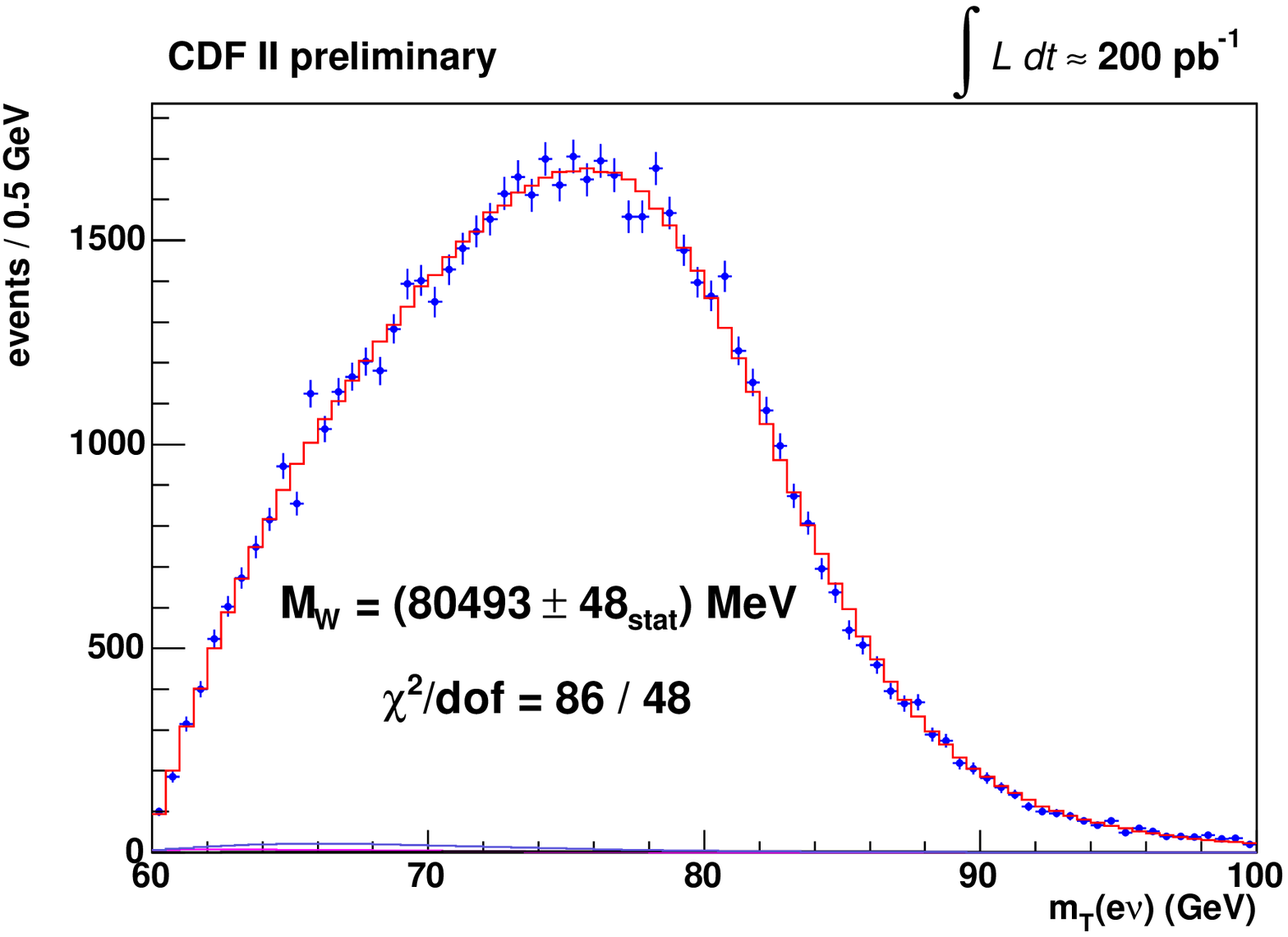}\epsfxsize=3.3in\epsfbox{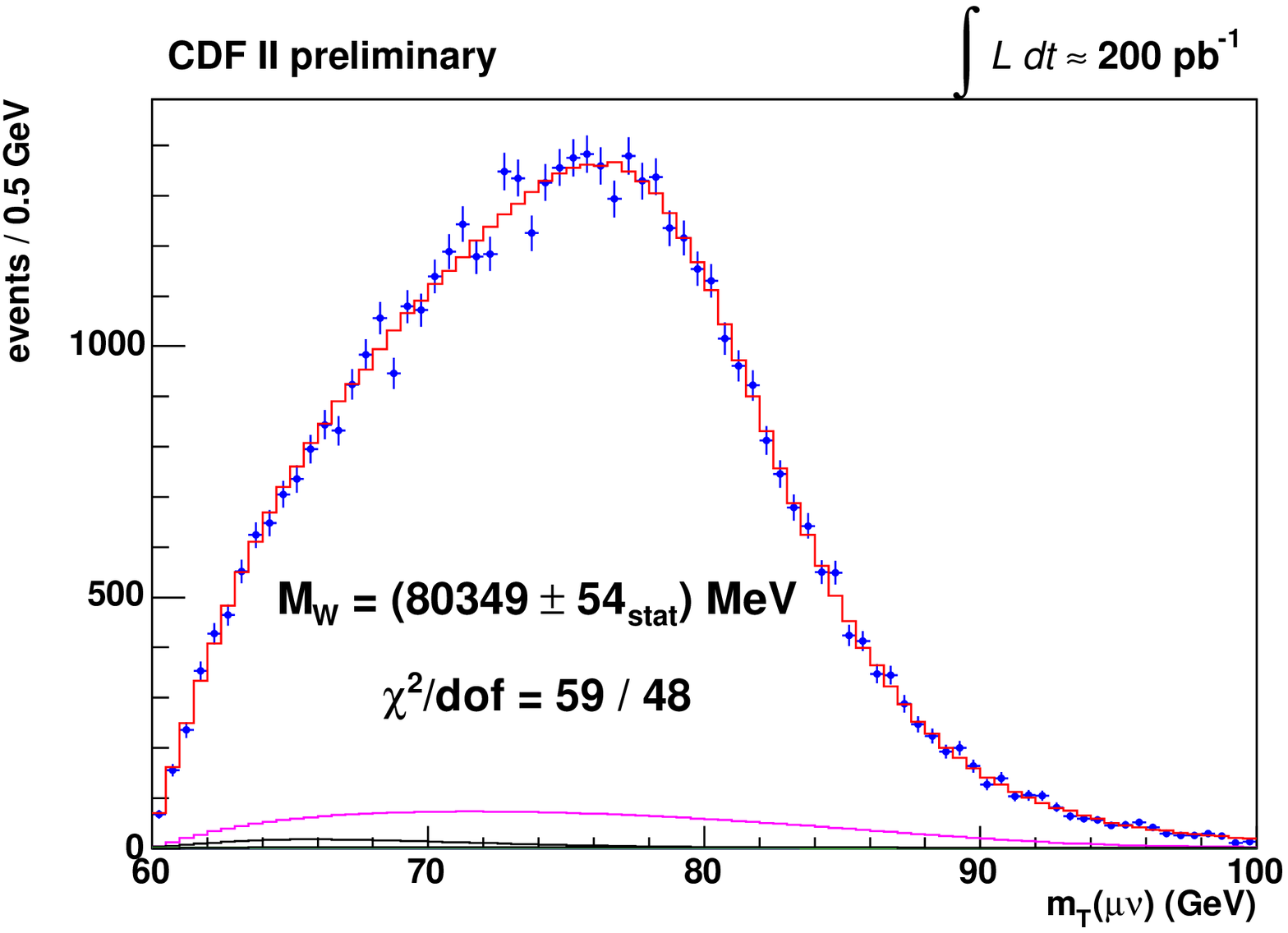}}
\caption{Left: Transverse mass fit in the electron decay channel. 
Right: Transverse mass fit in the muon decay channel. The histograms at the bottom of the
distributions represent the backgrounds. \label{results}}
\end{figure}

\begin{figure}[hb]
%=\epsfxsize=5cm   %width of figure - will enlarge/reduce the figures
%\epsfbox{fig3.eps}
%\figurebox{2cm}{3cm}{} %to have a box alone 
\centerline{\epsfxsize=3.8in\epsfbox{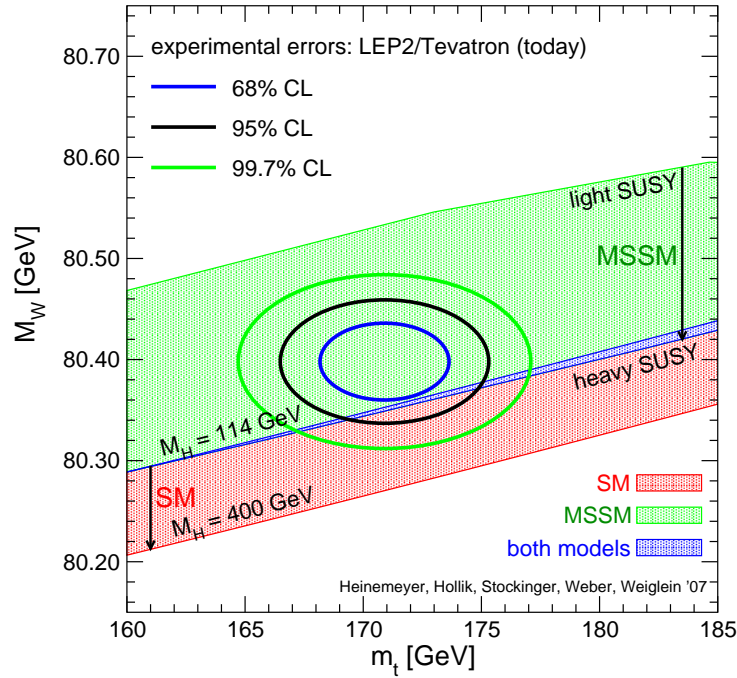}}
\caption{Constraint on the Higgs boson mass from direct $M_W$ and $M_{top}$ measurements 
at 68\%, 95\% and 99.7\% CL, along with SM and MSSM calculations. \label{higgs}}
\end{figure}

\begin{table}[ph]
\caption{Fit results from the distributions used to extract M$_W$ with uncertainties.}\label{fits} 
\vspace{0.4cm}
\begin{center}
\begin{tabular}{|c|cc|}
\hline
{} &{} &{}\\[-1.5ex]
Distribution & $W$ boson mass (MeV/$c^2$) & $\chi^2$/dof \\[1ex]
\hline
{} &{} &{} \\[-1.5ex]
$m_T(e,\nu)$ &80493$\pm$48$_{stat}$$\pm$39$_{syst}$ &86/48 \\[1ex]       
$p_T^l(e)$ &80451$\pm$58$_{stat}$$\pm$45$_{syst}$ &63/62 \\[1ex]     
$p_T^{\nu}(e)$ &80473$\pm$57$_{stat}$$\pm$54$_{syst}$ &63/62 \\[1ex] 
\hline
{} &{} &{} \\[-1.5ex]
$m_T(\mu,\nu)$ &80349$\pm$54$_{stat}$$\pm$27$_{syst}$ &59/48\\[1ex]
$p_T^l(\mu)$ &80321$\pm$66$_{stat}$$\pm$40$_{syst}$ &72/62\\[1ex]
$p_T^{\nu}(\mu)$ &80396$\pm$66$_{stat}$$\pm$46$_{syst}$ &44/62\\[1ex]
\hline
\end{tabular}
\end{center}
\end{table}
The transverse mass fit for the $W\rightarrow e\nu$ channel is shown in Fig. \ref{results} (left) and for the
$W\rightarrow\mu\nu$ channel in Fig. \ref{results} (right). 
The uncertainties for the $m_T$ fits in both channels are summarized in Table \ref{uncertainty}.
\begin{table}[ht]
\caption{Systematic and total uncertainties for the transverse mass fits. The last column shows the
correlated uncertainties, the last row is the quadrature sum of statistical and systematic uncertainty.}\label{uncertainty}
\vspace{0.4cm}
\begin{center}
\begin{tabular}{|l|c|c|c|}
\hline
{} &{} &{} &{} \\[-1.5ex]
Systematic (MeV/$c^2$) & $W\rightarrow e\nu$ & $W\rightarrow\mu\nu$ & Common \\[1ex]
\hline
{} &{} &{} &{}\\[-1.5ex]
Lepton Energy Scale& 30 & 17 & 17\\[1ex]
Lepton Energy Resolution& 9 & 3 & 0\\[1ex]
Recoil Energy Scale& 9 & 9 & 9\\[1ex]
Recoil Energy Resolution& 7 & 7 & 7\\[1ex]
Selection Bias & 3 & 1 & 0\\[1ex]
Lepton Removal & 8 & 5 & 5\\[1ex]
Backgrounds & 8 & 9 & 0\\[1ex]
$p_T(W)$ Model & 3 & 3 & 3\\[1ex]
Parton Distributions & 11 & 11 & 11\\[1ex]
QED radiation & 11 & 12 & 11\\[1ex]
\hline
{} &{} &{} &{}\\[-1.5ex]
Total Systematics & 39 & 27 & 26 \\[1ex]
\hline
{} &{} &{} &{}\\[-1.5ex]
Total Uncertainty & 62 & 60 & 26 \\[1ex]
\hline
\end{tabular}
\end{center}
\end{table}
We combine the six $W$ boson mass fits including all correlations to obtain
$M_W$=80413$\pm$34(stat)$\pm$34(syst) MeV/$c^2$. With a total uncertainty of
48 MeV/$c^2$, this measurement is the most precise single measurement to date. 
Inclusion of this result increases the world average
$W$ boson mass to $M_W$=80398$\pm$25 MeV/$c^2$ \cite{lepwmass}, reducing its uncertainty by 15\%. The updated world average impacts
the global precision electroweak fits, reducing the preferred Higgs boson mass fit
by 6 GeV/$c^2$ to $M_H$=76$^{+33}_{-24}$ GeV/$c^2$.
The resulting 95\% CL upper limit on the Higgs mass is 144 GeV/$c^2$ (182 GeV/$c^2$) with the
LEP II direct limit included (excluded) \cite{lepwmass}$^,$ \cite{higgslimit}. 
The direction of this change has interesting
theoretical implications: as Fig \ref{higgs} \cite{higgsconstraint} shows, the $M_W$ vs $M_{top}$ ellipse moves
a little deeper into the light-Higgs region excluded by LEP II and into the region 
favored by the minimal supersymmetry model (MSSM). 
While this is a one-sigma effect, it arouses further
interest in higher precision measurements of $M_W$ (and $M_{top}$).
Most of the systematic uncertainties in this measurement (Table \ref{uncertainty})
are limited by the statistics of the calibration samples used. Further improvements in the detector model
and the production and decay model (e.g. QED radiative corrections) are likely to shrink the other
systematic uncertainties as well. CDF has now accumulated an integrated luminosity of
about 2 fb$^{-1}$ and we look forward to a $W$ boson mass measurement with precision better
than the current world average of 25 MeV/$c^2$, with the dataset already in hand.

\section*{Acknowledgments}
I would like to thank my colleagues from the CDF collaboration
in particular the $W$ boson mass group for their hard work on 
this important analysis. Sincere thanks also to the conference 
organizers and participants for a superb conference. 
This work was supported by the European Commission under the
Marie Curie Programme.

\end{document}